\documentclass[12pt]{article}

\usepackage{graphicx}
\usepackage{amsmath}
\usepackage{amssymb}
\usepackage{amsthm}
\numberwithin{equation}{section}

\newcommand{\cS}{{\mathcal S}}
\newcommand{\cR}{{\mathcal R}}
\newcommand{\cN}{{\mathcal N}}
\newcommand{\cU}{{\mathcal U}}
\newcommand{\cZ}{{\mathcal Z}}
\newcommand{\fL}{{\mathfrak L}}
\newcommand{\fR}{{\mathfrak R}}
\newcommand{\fQ}{{\mathfrak Q}}
\newcommand{\fS}{{\mathfrak S}}
\newcommand{\fC}{{\mathfrak C}}
\newcommand{\fP}{{\mathfrak P}}
\newcommand{\fK}{{\mathfrak K}}
\newcommand{\fB}{{\mathfrak B}}
\newcommand{\fI}{{\mathfrak I}}
\newcommand{\fD}{{\mathfrak D}}
\newcommand{\fJ}{{\mathfrak J}}
\newcommand{\fF}{{\mathfrak F}}

\newcommand{\al}{\alpha}
\newcommand{\be}{\beta}
\newcommand{\ga}{\gamma}
\newcommand{\de}{\delta}
\newcommand{\ka}{\kappa}
\newcommand{\cT}{\mathcal T}
\newcommand{\bs}[1]{\boldsymbol{#1}}
\newcommand{\alg}[1]{\mathfrak{#1}}

\begin{document}

\baselineskip=16pt plus 0.2pt minus 0.1pt

\begin{titlepage}
\title{
\vspace{1cm}
{\Large\bf An Exceptional Algebraic Origin\\
of the AdS/CFT Yangian Symmetry}
}
\vskip20mm
\author{
{\sc Takuya Matsumoto}\thanks
{{\tt m05044c@math.nagoya-u.ac.jp}}\;
\quad and \quad
{\sc Sanefumi Moriyama}\thanks
{{\tt moriyama@math.nagoya-u.ac.jp}}\\[15pt]
{\it Graduate School of Mathematics, Nagoya University,} \\
{\it Nagoya 464-8602, Japan}
}
\date{\normalsize March, 2008}
\maketitle
\thispagestyle{empty}
\begin{abstract}
\normalsize
In the $\alg{su}(2|2)$ spin chain motivated by the AdS/CFT
 correspondence, a novel symmetry extending the superalgebra
 $\alg{su}(2|2)$ into $\alg{u}(2|2)$ was found.
We pursue the origin of this symmetry in the exceptional superalgebra
 $\alg{d}(2,1;\varepsilon)$, which recovers $\alg{su}(2|2)$ when the
 parameter $\varepsilon$ is taken to zero.
Especially, we rederive the Yangian symmetries of the AdS/CFT spin
 chain using the exceptional superalgebra and find that the
 $\varepsilon$-correction corresponds to the novel symmetry.
Also, we reproduce the non-canonical classical r-matrix of the AdS/CFT
 spin chain expressed with this symmetry from the canonical one of the
 exceptional algebra.
\end{abstract}
\end{titlepage}

\section{Introduction}
Symmetry algebra always plays an important role in physics.
The AdS/CFT correspondence is no exception, which relates the anomalous
dimension of operators in ${\cal N}=4$ super Yang-Mills theory with
Hamiltonian of string states on the ${\rm AdS}_5\times{\rm S}^5$
spacetime.
On the super Yang-Mills theory side, whose description is valid in the
weak {}'t Hooft coupling region $\lambda=g_{\rm YM}^2N\ll 1$, the
perturbative dilatation operator was mapped to the Hamiltonian of an
integrable spin chain model \cite{MZ} where infinite conserved charges
lead to solvability.
On the string sigma model side, valid in the strong coupling region
$\lambda\gg 1$, flat connection which implies infinite symmetries was
also constructed \cite{BPR}.
It was then expected that we can interpolate these two regions by
integrability and solve the theory completely.
See reviews \cite{review} for a summary of current progress and a list
of references.

This expectation was made more concrete in the subsequent analysis.
The $\cN=4$ super Yang-Mills theory has a global symmetry of the Lie
superalgebra $\alg{psu}(2,2|4)$, which is broken down to two copies of
$\alg{su}(2|2)$ algebra once we fix a vacuum.
The spin chain model with this $\alg{su}(2|2)$ symmetry, which
interpolates the two coupling regions, was constructed in
\cite{BSmatrix}.
To take care of the off-shell trace operators, the centrally extended
$\alg{su}(2|2)$ superalgebra was considered, which contains not only the
center of $\alg{su}(2|2)$, $\fC$, whose eigenvalue is the energy of the
spin chain, but also two additional centers, $\fP$ and $\fK$, whose
eigenvalues correspond to the momentum and should vanish on the on-shell
trace operators where the total momentum vanishes.
In other words, this spin chain has the
$\alg{psu}(2|2)\ltimes\mathbb{R}^3$ symmetry.
A remarkable property is that the S-matrix ${\mathcal S}$ on the
fundamental representation is determined uniquely by the Lie algebraic
symmetry up to an overall phase factor.
It was also checked that this S-matrix satisfies the Yang-Baxter
equation as the S-matrix of an integrable spin chain model is expected
to.

In order to clarify the full symmetry structure, it is an important
subject to study the universal R-matrix $\cS=\Pi\circ\cR$ that does not
depend on the representation.
Here $\Pi$ denotes the graded permutation operator.
The first step to investigate the universal R-matrix is to study its
classical limit $\cR_{12}=1+\hbar r_{12}+{\mathcal O}(\hbar^2)$.
It is known that a canonical form of the classical r-matrix
\begin{align}
r_{12}=\frac{\cT_{12}^\alg{g}}{i(u_1-u_2)}~,
\label{r}
\end{align}
is a solution to the classical Yang-Baxter equation, where
$\cT_{12}^\alg{g}$ is the two-site Casimir operator of the Lie algebra
$\alg{g}$ and $u$ is the spectral parameter.
This fact is the starting point for the classification of solutions of
the classical Yang-Baxter equation, as well as for the study of the
symmetry structure \cite{BD,M}.

In the case of the AdS/CFT spin chain, however, the relevant Killing
form of the centrally extended Lie superalgebra
$\alg{psu}(2|2)\ltimes{\mathbb R}^3$ is degenerate.
Hence, the quadratic Casimir operator $\cT^\alg{g}$ commuting with all
generators (and the corresponding two-site Casimir operator
$\cT^\alg{g}_{12}$) does not exist in the strict sense.
To cure the degeneracy of the Killing form, several different
regularizations were proposed.
The first one \cite{BSmatrix} is to consider the exceptional
superalgebra $\alg{d}(2,1;\varepsilon)$ which recovers
$\alg{psu}(2|2)\ltimes{\mathbb R}^3$ when the parameter $\varepsilon$ is
taken to zero.
In the exceptional algebra $\alg{d}(2,1;\varepsilon)$, the generators
$\fC$, $\fP$ and $\fK$ are not central but form the usual $\alg{su}(2)$
algebra.
The second one \cite{Banalytic} is to couple the central charges to the
$\alg{su}(2)$ outer automorphism, which has a similar origin as the
above case.
In \cite{MT}, to find an expression for the classical r-matrix similar
to \eqref{r}, another regularization was adopted by introducing a new
generator $\fI$ which complements the original superalgebra
$\alg{su}(2|2)$ into $\alg{u}(2|2)$ and couples to the center of
$\alg{su}(2|2)$.
Subsequently, the corresponding coproduct and antipode of this new
generator $\fI$ at level-1 were proposed \cite{MMT}
\begin{align}
\Delta\widehat\fI=\widehat\fI\otimes1+1\otimes\widehat\fI
+\frac{\hbar}{2}\left(\fQ^\al{}_a\cU^{-1}\otimes\fS^a{}_\al
+\fS^a{}_\al\cU^{+1}\otimes\fQ^\al{}_a\right)~,\quad
{\rm S}(\widehat\fI)=-\widehat\fI+2\hbar\fC~,
\label{coprod}
\end{align}
(with $\hbar$ related to the {}'t Hooft coupling $\lambda$ by
$1/\hbar=g=\sqrt{\lambda}/{4\pi}$)
and found to be a symmetry of the fundamental S-matrix.
However, it is still mysterious what the origin of this novel symmetry
is.

Though the original proposal of the classical r-matrix is complicated,
an elegant expression was found by \cite{BS}
\begin{align}
r_{12}&=\frac{\cT^{\alg{psu}}_{12}
-\cT^\alg{psu}\fD^{-1}\otimes\fD-\fD\otimes\cT^\alg{psu}\fD^{-1}}
{i(u_1-u_2)}\nonumber\\
&=\frac{\cT^{\alg{psu}}_{12}-(u_1/u_2)\cT^{\alg{psu}}\fC^{-1}\otimes\fC
-(u_2/u_1)\fC\otimes\cT^{\alg{psu}}\fC^{-1}}{i(u_1-u_2)}~,
\label{bs}
\end{align}
where $\cT^{\alg{psu}}$ is the quadratic Casimir-like operator of
$\alg{psu}(2|2)$ and $\cT^{\alg{psu}}_{12}$ is the corresponding
two-site operator.
In the above expression $\fD$ comes from the classical limit of the
centers $\fP$ and $\fK$ and relates to $\fC$ by $\fD=2u^{-1}\fC$.
The Casimir-like operator $\cT^{\alg{psu}}$ takes the opposite sign
on bosons and fermions
\begin{align}
\cT^{\alg{psu}}|\phi^a\rangle=-\frac{1}{4}|\phi^a\rangle~,\quad
\cT^{\alg{psu}}|\psi^\al\rangle=+\frac{1}{4}|\psi^\al\rangle~,
\end{align}
and serves the same role as $\fI$.
Interestingly, it was further noted that the eigenvalue of the new
generator $\fI$ and the composite operator $\cT^{\alg{psu}}\fC^{-1}$
coincides and that both the composite operator
$\cT^{\alg{psu}}\fC^{-1}$ and the Cartan generator of the
$\alg{su}(2)$ outer automorphism $2\fB=\fB^{1}{}_{1}-\fB^{2}{}_{2}$
satisfy similar commutation relations.

The above classical r-matrix \eqref{bs} implies a beautiful
Lie bialgebraic structure.
Especially, the classical cobrackets of all the generators, including
the new one $\fI$, were constructed.
This analysis indicates, among others, the Yangian symmetries of the
generator $\fI$ of all levels.
However, unfortunately, aside from the level-1 Yangian symmetry
\eqref{coprod}, we cannot show that they are exact quantum symmetries
of the fundamental S-matrix.
We believe that this unsatisfactory fact stems from the non-canonical
form of the classical r-matrix \eqref{bs}, or in other words, the
degeneracy of the Killing form of 
$\alg{psu}(2|2)\ltimes{\mathbb R}^3$.

In this paper we would like to pursue the origin of this novel generator
in the exceptional superalgebra $\alg{d}(2,1;\varepsilon)$ and interpret
the symmetry $\fI$ as the $\varepsilon$-correction of the
generators $\fC$, $\fP$ and $\fK$.
We first investigate the Yangian coproducts of $\fC$, $\fP$ and $\fK$ in
the exceptional superalgebra $\alg{d}(2,1;\varepsilon)$ and show that
the non-trivial coproduct of $\fI$ \eqref{coprod} coincides with the
$\varepsilon$-correction of the coproduct of $\fC$.
Secondly, we shall reproduce the classical r-matrix \eqref{bs} by taking
the $\varepsilon\to 0$ limit in the canonical classical r-matrix of the
exceptional superalgebra $\alg{d}(2,1;\varepsilon)$:
\begin{align}
r_{12}=\frac{\cT^{\alg{d}}_{12}}{i(u_1-u_2)}\Big|_{\varepsilon\to 0}~,
\label{eq:T^d_12}
\end{align}
where $T^{\alg{d}}_{12}$ is the two-site quadratic Casimir operator
of $\alg{d}(2,1;\varepsilon)$.
Since the generators $\fC$, $\fP$ and $\fK$ in
$\alg{d}(2,1;\varepsilon)$ are not central, the action on bosons and
fermions is slightly different.
As we shall see later, this is the origin of $\fI$ in the exceptional
algebra $\alg{d}(2,1;\varepsilon)$.

In the next section we first review the exceptional superalgebra
$\alg{d}(2,1;\varepsilon)$ and the $\alg{su}(2|2)$ spin chain model.
In section 3, We proceed to studying the Yangian coproducts of all the
$\alg{psu}(2|2)\ltimes{\mathbb R}^3$ generators from the superalgebra
$\alg{d}(2,1;\varepsilon)$ and find that the non-trivial coproduct of
the symmetry $\fI$ coincides with the $\varepsilon$-correction of that
of $\fC$.
Then we reproduce the non-canonical AdS/CFT classical r-matrix
from the canonical one of $\alg{d}(2,1;\varepsilon)$ in section 4.
Finally in section 5, we conclude with some discussions.

\section{Review of the exceptional Lie superalgebra}
In this section we would like to review the exceptional Lie superalgebra
$\alg{d}(2,1;\varepsilon)$, the limit which recovers the symmetry
$\alg{psu}(2|2)\ltimes{\mathbb R}^3$ and the AdS/CFT spin chain with
this symmetry \cite{BSmatrix}.

\subsection{Exceptional Superalgebra $\alg{d}(2,1;\varepsilon)$}
The exceptional Lie superalgebra $\alg{d}(2,1;\varepsilon)$ is given by
three orthogonal sets of $\alg{su}(2)$ triplet bosonic generators
$\fR^a{}_b$, $\fL^\al{}_\be$, $\fC^{\alg{a}}{}_{\alg{b}}$ and an octet
of fermionic generators $\fF^{a\al\alg{a}}$.
The non-trivial commutation relations between the $\alg{su}(2)$ bosonic
generators are given by
\begin{align}
[\fR^a{}_b,\fR^c{}_d]=\delta^c_b\fR^a{}_d-\delta^a_d\fR^c{}_b~,\quad
[\fL^\al{}_\be,\fL^\ga{}_\de]
=\delta^\ga_\be\fL^\al{}_\de-\delta^\al_\de\fL^\ga{}_\be~,\quad
[\fC^\alg{a}{}_\alg{b},\fC^\alg{c}{}_\alg{d}]
=\delta^\alg{c}_\alg{b}\fC^\alg{a}{}_\alg{d}
-\delta^\alg{a}_\alg{d}\fC^\alg{c}{}_\alg{b}~,
\end{align}
while the fermionic generators transform in the fundamental
representation of each $\alg{su}(2)$ as
\begin{align}
[\fR^a{}_b,\fF^{c\ga\alg{c}}]
=\delta^c_b\fF^{a\ga\alg{c}}
\!-\!\frac{1}{2}\delta^a_b\fF^{c\ga\alg{c}},\quad
[\fL^\al{}_\be,\fF^{c\ga\alg{c}}]
=\delta^\ga_\be\fF^{c\al\alg{c}}
\!-\!\frac{1}{2}\delta^\al_\be\fF^{c\ga\alg{c}},\quad
[\fC^\alg{a}{}_\alg{b},\fF^{c\ga\alg{c}}]
=\delta^\alg{c}_\alg{b}\fF^{c\ga\alg{a}}
\!-\!\frac{1}{2}\delta^\alg{a}_\alg{b}\fF^{c\ga\alg{c}}.
\end{align}
Also, the anti-commutation relation between the fermionic generators is
\begin{align}
\{\fF^{a\al\alg{a}},\fF^{b\be\alg{b}}\}
=\al\epsilon^{ak}\epsilon^{\al\be}\epsilon^{\alg{a}\alg{b}}
\fR^b{}_k
+\be\epsilon^{ab}\epsilon^{\al\ka}\epsilon^{\alg{a}\alg{b}}
\fL^\be{}_\ka
+\ga\epsilon^{ab}\epsilon^{\al\be}\epsilon^{\alg{a}\alg{k}}
\fC^\alg{b}{}_\alg{k}~,
\label{dJJ}
\end{align}
where the constants $\al$, $\be$, $\ga$ have to satisfy $\al+\be+\ga=0$
due to the Jacobi identity.
Since the overall rescaling does not change the algebraic structure, the
only one parameter which characterizes $\alg{d}(2,1;\varepsilon)$ is
$\varepsilon=-\ga/\al$.
The Killing form of this algebra is non-degenerate, and therefore, this
algebra has a well-defined quadratic Casimir operator,
\begin{align}
\cT^\alg{d}=\frac{1}{2}\Bigl(
-\al\alg{R}^a{}_b\alg{R}^b{}_a
-\be\alg{L}^\al{}_\be\alg{L}^\be{}_\al
-\ga\alg{C}^\alg{a}{}_\alg{b}\alg{C}^\alg{b}{}_\alg{a}
-\epsilon_{ab}\epsilon_{\al\be}\epsilon_{\alg{a}\alg{b}}
\fF^{a\al\alg{a}}\fF^{b\be\alg{b}}
\Bigr)~.
\label{dCasimir}
\end{align}

To reproduce the centrally extended superalgebra
$\alg{psu}(2|2)\ltimes{\mathbb R}^3$, let us choose
\begin{align}
\al=-1~,\quad\be=1-\varepsilon~,\quad\ga=\varepsilon~,
\end{align}
and take the limit $\varepsilon\to 0$.
To avoid the singular behavior and match the convention with the usual
one used in $\alg{psu}(2|2)\ltimes{\mathbb R}^3$, let us further rewrite
the last bosonic $\alg{su}(2)$ generator $\fC^\alg{a}{}_\alg{b}$ and the
fermionic generator $\fF^{a\alpha\alg{a}}$ as
\begin{align}
(\fC){}^\alg{a}{}_\alg{b}
=\frac{1}{\varepsilon}\begin{pmatrix}\fC&\fP\\-\fK&-\fC\end{pmatrix}~,\quad
(\fF^{a\al}){}^\alg{a}
=\begin{pmatrix}\epsilon^{ak}\fQ^\al{}_k\\
\epsilon^{\al\ka}\fS^a{}_\ka\end{pmatrix}~.
\end{align}
In the new convention, the anti-commutation relations between the
fermionic generators become
\begin{align}
\{\fQ^\al{}_a,\fQ^\be{}_b\}
&=\epsilon^{\al\be}\epsilon_{ab}\fP~,\nonumber\\
\{\fS^a{}_\al,\fS^b{}_\be\}
&=\epsilon^{ab}\epsilon_{\al\be}\fK~,\nonumber\\
\{\fQ^\al{}_a,\fS^b{}_\be\}
&=\delta^\al_\be\fR^b{}_a
+(1-\varepsilon)\delta^b_a\fL^\al{}_\be
+\delta^\al_\be\delta^b_a\fC~,
\label{CRsu}
\end{align}
while the commutation relations involving the last bosonic $\alg{su}(2)$
generators are
\begin{align}
&[\fC,\fP]=\varepsilon\fP~,\quad
[\fC,\fK]=-\varepsilon\fK~,\quad
[\fP,\fK]=-2\varepsilon\fC~,\nonumber\\
&[\fC,\fQ^\alpha{}_a]=+(\varepsilon/2)\fQ^\alpha{}_a~,\quad
[\fP,\fQ^\alpha{}_a]=0~,\quad
[\fK,\fQ^\alpha{}_a]
=\varepsilon\epsilon^{\alpha\beta}\epsilon_{ab}\fS^b{}_\beta~,
\nonumber\\
&[\fC,\fS^a{}_\alpha]=-(\varepsilon/2)\fS^a{}_\alpha~,\quad
[\fP,\fS^a{}_\alpha]
=-\varepsilon\epsilon^{ab}\epsilon_{\alpha\beta}\fQ^\beta{}_b~,\quad
[\fK,\fS^a{}_\alpha]=0~.
\label{RHSepsilon}
\end{align}
The rest commutation relations involving the other two $\alg{su}(2)$
generators $\fR^a{}_b$ and $\fL^\al{}_\be$ do not contain the parameter
$\varepsilon$.
These commutation relations are easily identified with those of
$\alg{psu}(2|2)\ltimes{\mathbb R}^3$ in the limit $\varepsilon\to 0$.
Note that, among others, $\fC$, $\fP$ and $\fK$ become central in this
limit.

The root lattice should be helpful in understanding the 
$\varepsilon\to 0$ limit intuitively. (See Fig.~1.)
As we will see in subsection 4.2, the additional dimension of the
$\alg{d}(2,1;\varepsilon)$ root lattice plays an important role in
constructing the representation of $\alg{d}(2,1;\varepsilon)$.
In the next subsection let us recapitulate the representation of 
$\alg{psu}(2|2)\ltimes{\mathbb R}^3$ first.

\begin{figure}[t]
\begin{center}
\scalebox{1.0}[1.0]{\includegraphics{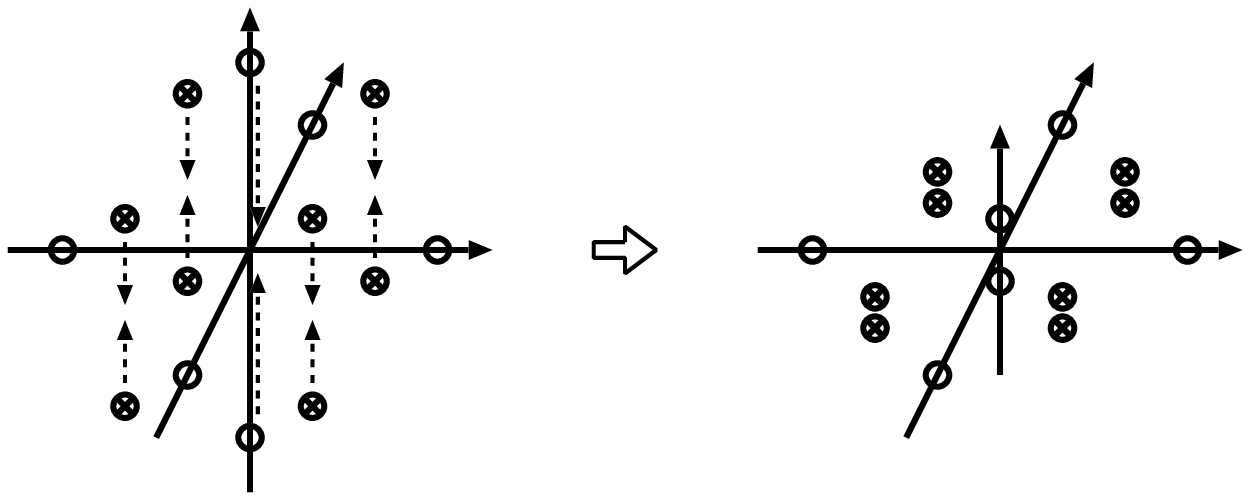}}
\setlength{\unitlength}{1mm}
\begin{picture}(20,20)
\put(-4,20){\makebox(10,10){$\fR$}}
\put(-10,16){\makebox(10,10){$\fR^1{}_2$}}
\put(-56,24){\makebox(10,10){$\fR^2{}_1$}}
\put(-22,42){\makebox(10,10){$\fL$}}
\put(-28,38){\makebox(10,10){$\fL^1{}_2$}}
\put(-36,4){\makebox(10,10){$\fL^2{}_1$}}
\put(-34,36){\makebox(10,10){$\fC$}}
\put(-28,24){\makebox(10,10){$\fP$}}
\put(-36,18){\makebox(10,10){$\fK$}}
\put(-14,32){\makebox(10,10){$\fQ^1{}_2$}}
\put(-14,28){\makebox(10,10){$\fS^1{}_2$}}
\put(-20,14){\makebox(10,10){$\fQ^2{}_2$}}
\put(-20,10){\makebox(10,10){$\fS^1{}_1$}}
\put(-44,32){\makebox(10,10){$\fQ^1{}_1$}}
\put(-44,28){\makebox(10,10){$\fS^2{}_2$}}
\put(-50,14){\makebox(10,10){$\fQ^2{}_1$}}
\put(-50,10){\makebox(10,10){$\fS^2{}_1$}}
\put(-109,48){\makebox(10,10){$\fC$}}
\put(-81,20){\makebox(10,10){$\fR$}}
\put(-99,42){\makebox(10,10){$\fL$}}
\put(-70,16){\makebox(10,10){$\varepsilon\to 0$}}
\end{picture}
\end{center}
\caption{Root lattice of $\alg{d}(2,1;\varepsilon)$ and the limit to
reproduce $\alg{psu}(2|2)\ltimes{\mathbb R}^3$.
Here we have defined $\fR=(\fR^1{}_1-\fR^2{}_2)/2$ and
 $\fL=(\fL^1{}_1-\fL^2{}_2)/2$.}
\end{figure}

\subsection{Fundamental representation of
$\alg{psu}(2|2)\ltimes{\mathbb R}^3$}
The algebra $\alg{psu}(2|2)\ltimes{\mathbb R}^3$ has a
$\bs{2}|\bs{2}$-dimensional representation, which is called the
fundamental representation.
The states in the representation space are labeled by two bosons 
$|\phi^a\rangle$ and two fermions $|\psi^\al\rangle$.

This representation has a physical meaning in the $\alg{su}(2|2)$ spin
chain model motivated by super Yang-Mills theory.
In the model, the vacuum state is identified with the infinitely long
trace operator
\begin{align}
|0\rangle=|\cZ\cZ\cdots\cZ\rangle
\quad\Leftrightarrow\quad{\rm Tr}(\cZ\cZ\cdots\cZ)~,
\end{align} 
where $\cZ$ corresponds to one of the complex scalar fields in super
Yang-Mills theory.
This identification of the vacuum state breaks the original global
$\alg{psu}(2,2|4)$ symmetry into two copies of the $\alg{su}(2|2)$
symmetry.
Hence, the excitation of the original super Yang-Mills theory is
expressed by two excitations in each $\alg{su}(2|2)$ spin chain.
The excitation $\chi\in\{\phi^1,\phi^2|\psi^1,\psi^2\}$ in the
$\alg{su}(2|2)$ spin chain form the fundamental representation of
$\alg{su}(2|2)$.
We shall consider the $K$-magnon asymptotic state where the excitations
are well-separated:
\begin{align}
|\chi_1\chi'_2\cdots\chi''_K\rangle
=\sum_{n_1\ll n_2\ll\cdots\ll n_K}e^{ip_1n_1}e^{ip_2n_2}\cdots e^{ip_Kn_K}
|\cZ\cdots\cZ\chi_1\cZ\cdots\cZ\chi'_2\cZ
\cdots\cZ\chi''_K\cZ\cdots\cZ\rangle~,
\label{asympt}
\end{align}
with $n_k$ denoting the site of the $k$-th excitation $\chi_k$.
Here, the excitation $\chi_k$ carries the momentum $p_k$ on the spin
chain.

\begin{figure}[t]
\begin{center}
\scalebox{1.0}[1.0]{\includegraphics{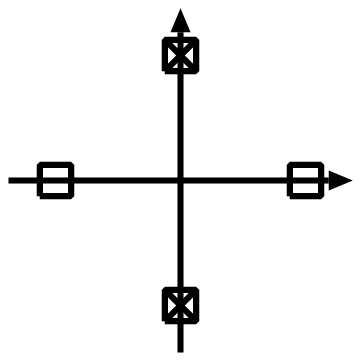}}
\setlength{\unitlength}{1mm}
\begin{picture}(20,20)
\put(-6,12){\makebox(10,10){$\fR$}}
\put(-24,32){\makebox(10,10){$\fL$}}
\put(-12,8){\makebox(10,10){$|\phi^1\rangle$}}
\put(-40,8){\makebox(10,10){$|\phi^2\rangle$}}
\put(-20,24){\makebox(10,10){$|\psi^1\rangle$}}
\put(-20,0){\makebox(10,10){$|\psi^2\rangle$}}
\end{picture}
\end{center}
\caption{Weight lattice of the fundamental representation of
$\alg{psu}(2|2)$.}
\end{figure}

The two sets of $\alg{su}(2)$ bosonic generators $\fR^a{}_b$ and
$\fL^\al{}_\be$ act kinematically on the states:
\begin{align}
\fR^a{}_b|\phi^c_k\rangle
=\de^c_b|\phi^a_k\rangle-\dfrac{1}{2}\de^a_b|\phi^c_k\rangle~,\quad
\fL^\al{}_\be|\psi^\ga_k\rangle
=\de^\ga_\be|\psi^\al_k\rangle-\dfrac{1}{2}\de^\al_\be|\psi^\ga_k\rangle~. 
\end{align}
Let us postulate the generic action of the fermionic generators
$\fQ^\al{}_a$ and $\fS^a{}_\al$ as
\begin{align}
&{\fQ}^\alpha{}_a|\phi^b_k\rangle
=a_k{\,}\delta^b_a
|\cZ^{+1/2}\psi^\alpha_k\rangle~,\quad
{\fQ}^\alpha{}_a|\psi^\beta_k\rangle
=b_k{\,}\epsilon^{\alpha\beta}\epsilon_{ab}
|\cZ^{+1/2}\phi^b_k\rangle~,\nonumber\\
&{\fS}^a{}_\alpha|\phi^b_k\rangle
=c_k{\,}\epsilon^{ab}\epsilon_{\alpha\beta}
|\cZ^{-1/2}\psi^\beta_k\rangle~,\quad
{\fS}^a{}_\alpha|\psi^\beta_k\rangle
=d_k{\,}\delta^\beta_\alpha
|\cZ^{-1/2}\phi^a_k\rangle~.
\label{QSphipsi}
\end{align}
(See Fig.~2.)
Here $\cZ^{\pm 1(/2)}$ stands for insertion or removal of (half of) the
$\cZ$ field.
This is important when we consider the action on the multi-magnon
state.
If we stick to the convention of placing the extra $\cZ$ fields on the
most left, a braiding factor $U_k=e^{ip_k/2}$ will emerge in moving the
$\cZ$ field to the left as in
$|\chi_k\cZ^n\rangle=U_k^{2n}|\cZ^n\chi_k\rangle$.
This effect can alternatively be interpreted as the inclusion of a
braiding factor $U_k=e^{ip_k/2}$ in the tensor product such as
coproducts and two-site Casimir operators \cite{Hopf}.
The power of the $\cZ$ field $n$, which depends on the generators, is
called grading and corresponds to the eigenvalues of $\fC$ in the
adjoint representation \eqref{RHSepsilon} (in the unit of
$\varepsilon$).

The consistency condition with the algebra, especially \eqref{CRsu} in
the limit $\varepsilon\to 0$, requires the coefficients $a_k$, $b_k$,
$c_k$ and $d_k$ to satisfy the relation
\begin{align}
a_kd_k-b_kc_k=1~,
\label{ad-bc}
\end{align}
as well as determines the action of centers $\fC$, $\fP$ and $\fK$
\begin{align}
\fC|\chi_k\rangle=\dfrac{1}{2}(a_kd_k+b_kc_k)|\chi_k\rangle~,\quad
\fP|\chi_k\rangle=a_kb_k|{\mathcal Z}^{+1}\chi_k\rangle~,\quad
\fK|\chi_k\rangle=c_kd_k|{\mathcal Z}^{-1}\chi_k\rangle~,
\end{align}
where $\chi_k$ is an arbitrary $k$-th excitation.
The coefficients $a_k$, $b_k$, $c_k$ and $d_k$ are expressed by
parameters $x^\pm_k$, $\ga_k$ and a constant $\al$:
\begin{align}
a_k=\sqrt{g}\ga_k~,\quad
b_k=\sqrt{g}\frac{\al}{\ga_k}\left(1-\frac{x^+_k}{x^-_k}\right)~,\quad
c_k=\sqrt{g}\frac{i\ga_k}{\al x^+_k}~,\quad
d_k=\sqrt{g}\frac{x^+_k}{i\ga_k}\left(1-\frac{x^-_k}{x^+_k}\right)~,
\end{align}
where the parameters $x^\pm_k$ relate to the momentum of the $k$-th
excitation by
\begin{align}
\frac{x^+_k}{x^-_k}=e^{ip_k}(=U_k^2)~,
\label{U}
\end{align}
and obey the constraint
\begin{align}
x^+_k+\frac{1}{x^+_k}-x^-_k-\frac{1}{x^-_k}=\frac{i}{g}~.
\label{xpm}
\end{align}
Note that we have required the extra central charges $\fP$ and $\fK$ to
vanish on the on-shell states where the total momentum vanishes.
This is why the parameter $\al$ has to be a constant independent of the
excitations.

The constraint \eqref{xpm} can be solved explicitly \cite{AF}
\begin{align}
x^\pm_k=x_k\biggl(\sqrt{1-\frac{1}{[2g(x_k-x_k^{-1})]^2}}
\pm\frac{i}{2g(x_k-x_k^{-1})}\biggr)~.
\label{xpmsol}
\end{align}
In the classical limit $g\to\infty$, the coefficients $a_k$, $b_k$,
$c_k$ and $d_k$ can be expressed as \cite{T}
\begin{align}
a_k=\sqrt{g}\gamma_k~,\quad
b_k=\frac{\alpha}{i\sqrt{g}\gamma_k}\frac{1}{x_k-x_k^{-1}}~,\quad
c_k=\frac{i\sqrt{g}\gamma_k}{\alpha}x_k^{-1}~,\quad
d_k=\frac{1}{\sqrt{g}\gamma_k}\frac{x_k}{x_k-x_k^{-1}}~.
\label{abcd}
\end{align}
To keep all the variables finite in the classical limit, we assume that
$\gamma_k$ scales as $1/\sqrt{g}$.
Note that in the classical limit, the braiding factor reduces to
$U_k=1$.

\section{Yangian coproducts from $\alg{d}(2,1;\varepsilon)$} 
After reviewing the superalgebra $\alg{d}(2,1;\varepsilon)$ and the limit
leading to the symmetry $\alg{psu}(2|2)\ltimes{\mathbb R}^3$, let us
turn to the subject of this paper.
We shall reproduce the coproducts of the generators in 
$\alg{psu}(2|2)\ltimes{\mathbb R}^3$ from those in
$\alg{d}(2,1;\varepsilon)$.
In the derivation, we will see the first sign that the novel generator
$\fI$ \cite{MT,MMT,BS} can be interpreted as the
$\varepsilon$-correction of the last $\alg{su}(2)$ generators
$\fC^\alg{a}{}_\alg{b}$.

A model is solvable if it has an equal number of conserved charges and
degrees of freedom.
For a solvable field-theoretical model with infinite degrees of freedom,
we expect infinite conserved charges.
In the field-theoretical model the scattering process between incoming
and outgoing states is described by the S-matrix.
For the solvable model, it is often the case that multi-body S-matrix
factorizes into the product of two-body S-matrix.
To see the symmetry of the S-matrix, we have to specify how the
symmetries $\fJ^A$ act on two-body states, which can be non-local.
The mathematical words for this action is called coproduct
$\Delta\fJ^A$.
The R-matrix ${\mathcal R}=\Pi\circ{\mathcal S}$ of the rational type
has a symmetry called the Yangian algebra $[\Delta\widehat\fJ^A,\cS]=0$,
besides the Lie algebraic symmetries $[\Delta\fJ^A,\cS]=0$.
In general, the Yangian algebra $Y(\alg{g})$, associated with Lie
algebra $\alg{g}$, is a kind of the universal enveloping algebra
$U(\alg{g}[u,u^{-1}])$ of the loop algebra $\alg{g}[u,u^{-1}]$.

In \cite{BYangian}, the level-1 Yangian symmetry $\widehat\fJ^A=iu\fJ^A$
(on the evaluation representation) associated with
$\alg{psu}(2|2)\ltimes{\mathbb R}^3$ is obtained from the standard
formula \cite{M}
\begin{align}
\Delta\widehat\fJ^A=\widehat\fJ^A\otimes 1+\cU^{[A]}\otimes\widehat\fJ^A
+\dfrac{\hbar}{2}f^A_{BC}\fJ^B\cU^{[C]}\otimes\fJ^C~,\quad
{\rm S}(\widehat\fJ^A)=-\cU^{-[A]}\Bigl(\widehat\fJ^A
-\dfrac{\hbar}{4}f^A_{BC}f^{BC}_D\fJ^D\Bigr)~, 
\end{align}
where $[A]$ is the grading charge and the abelian generator $\cU$ is the
braiding factor \eqref{U}.
The first formula of the coproduct can be reexpressed as
\begin{align}
\Delta\widehat\fJ^A=\widehat\fJ^A\otimes 1+\cU^{[A]}\otimes\widehat\fJ^A
+\dfrac{\hbar}{4}
[\cT^{\alg{g}}_{12},\cU^{[A]}\otimes\fJ^A-\fJ^A\otimes 1]~,
\label{coprd}
\end{align}
which is very useful in calculation.

The two-site Casimir operator of the superalgebra
$\alg{d}(2,1;\varepsilon)$ can be read off directly from
\eqref{dCasimir} as
\begin{align}
\cT^{\alg{d}}_{12}
=\fR^a{}_b\otimes\fR^b{}_a
-(1-\varepsilon)\fL^\al{}_\be\otimes\fL^\be{}_\al
+\fQ^\al{}_b\cU^{-1}\otimes\fS^b{}_\al
-\fS^a{}_\be\cU^{+1}\otimes\fQ^\be{}_a
-\frac{1}{\varepsilon}\cT^\alg{C}_{12}~,
\label{dtwosite}
\end{align}
with $\cT^\alg{C}_{12}$ being the Casimir operator of the last
$\alg{su}(2)$ generators
\begin{align}
\cT^\alg{C}_{12}
=-\fP\cU^{-2}\otimes\fK+2\fC\otimes\fC-\fK\cU^{+2}\otimes\fP~.
\label{CTcasimir}
\end{align}
Using this two-site Casimir operator \eqref{dtwosite}, we find without
difficulty that
\begin{align}
&\Delta\widehat\fR^a{}_b
=\widehat\fR^a{}_b\otimes 1+1\otimes\widehat\fR^a{}_b
+\frac{\hbar}{2}\big[\fR^a{}_c\otimes\fR^c{}_b
-\fR^c{}_b\otimes\fR^a{}_c
\nonumber\\
&\quad-\fS^a{}_\ga\cU\otimes\fQ^\ga{}_b
-\fQ^\ga{}_b\cU^{-1}\otimes\fS^a{}_\ga
+\frac{1}{2}\delta^a_b\big(\fS^c{}_\ga\cU\otimes\fQ^\ga{}_c
+\fQ^\ga{}_c\cU^{-1}\otimes\fS^c{}_\ga\big)\big]~,\nonumber\\
&\Delta\widehat\fL^\al{}_\be
=\widehat\fL^\al{}_\be\otimes 1+1\otimes\widehat\fL^\al{}_\be
+\frac{\hbar}{2}\big[-(1-\varepsilon)
\big(\fL^\al{}_\ga\otimes\fL^\ga{}_\be
-\fL^\ga{}_\be\otimes\fL^\al{}_\ga\big)
\nonumber\\
&\quad+\fQ^\al{}_c\cU^{-1}\otimes\fS^c{}_\be
+\fS^c{}_\be\cU\otimes\fQ^\al{}_c
-\frac{1}{2}\delta^\al_\be\big(\fQ^\ga{}_c\cU^{-1}\otimes\fS^c{}_\ga
+\fS^c{}_\ga\cU\otimes\fQ^\ga{}_c\big)\big]~,\nonumber\\
&\Delta\widehat\fQ^\al{}_a
=\widehat\fQ^\al{}_a\otimes 1+\cU\otimes\widehat\fQ^\al{}_a 
+\frac{\hbar}{2}\big[
\epsilon^{\al\be}\epsilon_{ab}\left(\fP\cU^{-1}\otimes\fS^b{}_\be
-\fS^b{}_\be\cU^2\otimes\fP\right)\nonumber\\
&\quad-\left(\de^\al_\be\fR^b{}_a
+(1-\varepsilon)\de^b_a\fL^\al{}_\be
+\de^b_a\de^\al_\be\fC\right)\cU\otimes\fQ^\be{}_b
+\fQ^\be{}_b\otimes\left(\de^\al_\be\fR^b{}_a
+(1-\varepsilon)\de^b_a\fL^\al{}_\be+\de^b_a\de^\al_\be\fC\right) 
\big]~,\nonumber\\
&\Delta\widehat\fS^a{}_\al
=\widehat\fS^a{}_\al\otimes 1+\cU^{-1}\otimes\widehat\fS^a{}_\al 
+\frac{\hbar}{2}\big[
-\epsilon^{ab}\epsilon_{\al\be}\left(\fK\cU\otimes\fQ^\be{}_b
-\fQ^\be{}_b\cU^{-2}\otimes\fK\right)\nonumber\\
&\quad+\left(\de^\be_\al\fR^a{}_b
+(1-\varepsilon)\de^a_b\fL^\be{}_\al
+\de^a_b\de^\be_\al\fC\right)\cU^{-1}\otimes\fS^b{}_\be
-\fS^b{}_\be\otimes\left(\de^\be_\al\fR^a{}_b
+(1-\varepsilon)\de^a_b\fL^\be{}_\al+\de^a_b\de^\be_\al\fC\right) 
\big]~,\nonumber\\
&\Delta\widehat\fC=\widehat\fC\otimes 1+1\otimes\widehat\fC 
+\frac{\hbar}{2}\big[\fP\cU^{-2}\otimes\fK-\fK\cU^2\otimes\fP 
+\frac{\varepsilon}{2}\left(\fQ^\al{}_a\cU^{-1}\otimes\fS^a{}_\al
+\fS^a{}_\al\cU\otimes\fQ^\al{}_a\right)\big]~,\nonumber\\
&\Delta\widehat\fP=\widehat\fP\otimes 1+\cU^2\otimes\widehat\fP
+\hbar\big[-\fC\cU^2\otimes\fP+\fP\otimes\fC
+\frac{\varepsilon}{2}\epsilon^{ab}\epsilon_{\al\be}
\fQ^\al{}_a\cU\otimes\fQ^\be{}_b\big]~,\nonumber\\
&\Delta\widehat\fK=\widehat\fK\otimes 1+\cU^{-2}\otimes\widehat\fK
+\hbar\big[\fC\cU^{-2}\otimes\fK-\fK\otimes\fC
+\frac{\varepsilon}{2}\epsilon^{\al\be}\epsilon_{ab}
\fS^a{}_\al\cU^{-1}\otimes\fS^b{}_\be\big]~.
\label{Ccoprod}
\end{align}
The antipode is also recovered,
${\rm S}(\widehat\fJ^A)=-\cU^{-[A]}\widehat\fJ^A$, if we assume the
counit vanishes, $\epsilon(\widehat\fJ^A)=0$.

When we plug the Casimir operator of $\alg{d}(2,1;\varepsilon)$
\eqref{dtwosite} into the coproduct formula \eqref{coprd}, the last term
in \eqref{dtwosite} seems divergent in the limit $\varepsilon\to 0$ at
the first sight.
However, the divergence is canceled by $\varepsilon$ on the
right-hand-side of \eqref{RHSepsilon}, which makes $\fC$, $\fP$ and
$\fK$ central in the limit $\varepsilon\to 0$.
Now it is easy to see that we recover the previous coproducts in
\cite{BYangian} after taking the limit $\varepsilon\to 0$.
Note that the $\varepsilon$-correction of $\Delta\widehat\fC$ in
\eqref{Ccoprod} is exactly the non-trivial part of the symmetry
$\Delta\widehat\fI$ \eqref{coprod}.
Hence, it is natural to regard the secret symmetry $\fI$ as the
$\varepsilon$-correction of the generator $\fC$.

In \cite{BYangian} the coproduct of $\alg{psu}(2|2)\ltimes{\mathbb R}^3$
was found from the $\alg{su}(2)$ outer automorphism.
There is an interesting way to reproduce similar commutation relations
and Casimir operator used in the derivation with the $\alg{su}(2)$
outer automorphism.
Let us separate the last $\alg{su}(2)$ generators
$\fC^\alg{a}{}_\alg{b}$ of $\alg{d}(2,1;\varepsilon)$ into
\begin{align}
\fC^\alg{a}{}_\alg{b}
=\frac{1}{\varepsilon}\overline\fC{}^\alg{a}{}_\alg{b}
+\fB^\alg{a}{}_\alg{b}~,
\end{align}
where $\overline\fC{}^\alg{a}{}_\alg{b}$ is interpreted as the center
of $\alg{psu}(2|2)\ltimes{\mathbb R}^3$ and $\fB^\alg{a}{}_\alg{b}$ is 
their difference.
Then, the whole commutation relations involving $\fC^\alg{a}{}_\alg{b}$ 
can be reproduced by the commutation relations\footnote{Note that the 
second equation is not the commutation relation of the usual Lie
algebra.
Therefore, the generators $\overline\fC{}^\alg{a}{}_\alg{b}$ and
$\fB^\alg{a}{}_\alg{b}$ do not satisfy all the Jacobi identities.
We are grateful to N.~Beisert for pointing this out.} at each order of
$\varepsilon$:
\begin{align}
&[\overline\fC{}^\alg{a}{}_\alg{b},
\overline\fC{}^\alg{c}{}_\alg{d}]=0~,\quad
[\overline\fC{}^\alg{a}{}_\alg{b},\fB^\alg{c}{}_\alg{d}]
+[\fB^\alg{a}{}_\alg{b},\overline\fC{}^\alg{c}{}_\alg{d}]
=\de^\alg{c}_\alg{b}\overline\fC{}^\alg{a}{}_\alg{d}
-\de^\alg{a}_\alg{d}\overline\fC{}^\alg{c}{}_\alg{b}~,\quad
[\fB^\alg{a}{}_\alg{b},\fB^\alg{c}{}_\alg{d}]
=\de^\alg{c}_\alg{b}\fB^\alg{a}{}_\alg{d}
-\de^\alg{a}_\alg{d}\fB^\alg{c}{}_\alg{b}~,\nonumber\\
&[\overline\fC{}^\alg{a}{}_\alg{b},\fF^{c\ga\alg{c}}]=0~,\quad
[\fB^\alg{a}{}_\alg{b},\fF^{c\ga\alg{c}}]
=\de^\alg{c}_\alg{b}\fF^{c\ga\alg{a}}
-\dfrac{1}{2}\de^\alg{a}_\alg{b}\fF^{c\ga\alg{c}}~.
\label{B}
\end{align}
and the two-site Casimir operator \eqref{dtwosite} reduces to
\begin{align}
&\cT^\alg{d}_{12}
=\fR^a{}_b\otimes\fR^b{}_a-\fL^\alpha{}_\beta\otimes\fL^\beta{}_\alpha
+\fQ^\alpha{}_b\otimes\fS^b{}_\alpha
-\fS^a{}_\beta\otimes\fQ^\beta{}_a\nonumber\\
&\qquad-\dfrac{1}{\varepsilon}
\overline\fC{}^\alg{a}{}_\alg{b}\otimes\overline\fC{}^\alg{b}{}_\alg{a}
-\overline\fC{}^\alg{a}{}_\alg{b}\otimes\fB^\alg{b}{}_\alg{a}
-\fB^\alg{a}{}_\alg{b}\otimes\overline\fC{}^\alg{b}{}_\alg{a} 
+{\mathcal O}(\varepsilon)~,
\label{BCasimir}
\end{align}
(where we have dropped all the braiding factors $\cU$ for simplicity).
Note that the second equation of \eqref{B} is not conventional and in
the two-site Casimir operator \eqref{BCasimir} we have an extra term
of $(1/\varepsilon)
\overline\fC{}^\alg{a}{}_\alg{b}\otimes\overline\fC{}^\alg{b}{}_\alg{a}$
compared with \cite{BYangian}.
The effect of these two differences adds up to the correct
coefficients of \cite{BYangian}.

\section{Non-canonical classical r-matrix from
$\alg{d}(2,1;\varepsilon)$}
In the previous section, we have seen that the coproducts of the Yangian
generators can be reproduced from the exceptional superalgebra
$\alg{d}(2,1;\varepsilon)$, which implies that the AdS/CFT spin chain has
an origin in the exceptional superalgebra $\alg{d}(2,1;\varepsilon)$ and
the symmetry $\fI$ can be regarded as the $\varepsilon$-correction
of the last $\alg{su}(2)$ generators $\fC^\alg{a}{}_\alg{b}$.
Here, we would like to reproduce the non-canonical AdS/CFT classical
r-matrix \eqref{bs} from the canonical classical r-matrix of this
exceptional superalgebra.

\subsection{Observation} 
The classical r-matrix of the $\alg{su}(2|2)$ spin chain was obtained in
\cite{MT,BS}
\begin{align}
r_{12}=\frac{\cT^{\alg{psu}}_{12}
-\cT^\alg{psu}\fD^{-1}\otimes\fD-\fD\otimes\cT^\alg{psu}\fD^{-1}}
{i(u_1-u_2)}~,
\label{bsr}
\end{align}
where $\cT^{\alg{psu}}$ is the Casimir-like operator of
$\alg{psu}(2|2)$ and $\cT^{\alg{psu}}_{12}$ is the corresponding
two-site operator:
\begin{align}
\cT^{\alg{psu}}
&=\dfrac{1}{2}\left(\fR^a{}_b\fR^b{}_a
-\fL^\alpha{}_\beta\fL^\beta{}_\alpha
+\fQ^\alpha{}_b\fS^b{}_\alpha
-\fS^a{}_\beta\fQ^\beta{}_a\right)~,
\nonumber\\
\cT^{\alg{psu}}_{12}
&=\fR^a{}_b\otimes\fR^b{}_a
-\fL^\alpha{}_\beta\otimes\fL^\beta{}_\alpha
+\fQ^\alpha{}_b\cU^{-1}\otimes\fS^b{}_\alpha
-\fS^a{}_\beta\cU^{+1}\otimes\fQ^\beta{}_a~.
\end{align}
This classical r-matrix takes, however, a non-canonical form, where the
numerator of \eqref{bsr} is not the Casimir operator of the symmetry
algebra.
Our goal in this section is to derive this non-canonical classical
r-matrix from the canonical r-matrix of the exceptional algebra
$\alg{d}(2,1;\varepsilon)$ by taking $\varepsilon$ to zero:
\begin{align}
r_{12}=\frac{\cT^{\alg{d}}_{12}}{i(u_1-u_2)}\Big|_{\varepsilon\to 0}~,
\end{align}
where $\cT^{\alg{d}}_{12}$ is the two-site Casimir operator
\eqref{dtwosite} of $\alg{d}(2,1;\varepsilon)$,
\begin{align}
\cT^{\alg{d}}_{12}
=\fR^a{}_b\otimes\fR^b{}_a
-(1-\varepsilon)\fL^\al{}_\be\otimes\fL^\be{}_\al
+\fQ^\al{}_b\cU^{-1}\otimes\fS^b{}_\al
-\fS^a{}_\be\cU^{+1}\otimes\fQ^\be{}_a
-\frac{1}{\varepsilon}\cT^\alg{C}_{12}~,
\label{d12}
\end{align}
with the last $\alg{su}(2)$ part being
$\cT^\alg{C}_{12}=-\fP\cU^{-2}\otimes\fK
+2\fC\otimes\fC-\fK\cU^{+2}\otimes\fP$.

The good property of the Casimir-like operator $\cT^{\alg{psu}}$, which
appears in the non-canonical r-matrix \eqref{bsr}, is that it takes the
opposite sign on bosons and fermions,
\begin{align}
\cT^{\alg{psu}}|\phi^a\rangle=-\frac{1}{4}|\phi^a\rangle~,\quad
\cT^{\alg{psu}}|\psi^\alpha\rangle=+\frac{1}{4}|\psi^\alpha\rangle~,
\end{align}
and plays the role of the generator $\fI$.
This property is realized in the exceptional superalgebra
$\alg{d}(2,1;\varepsilon)$ as the difference of the action of
$\fC^\alg{a}{}_\alg{b}$ on bosons and fermions, since the generators
$\fC$, $\fP$ and $\fK$ are no longer central in
$\alg{d}(2,1;\varepsilon)$.
The precise meaning of this will be clarified when we consider the
representation in the following subsection.

Note that, although it is easy to see that the first four terms of
$T^{\alg{d}}_{12}$ in \eqref{d12} reduces to $\cT^{\alg{psu}}_{12}$ in the
limit $\varepsilon\to 0$, the final term
$\cT^\alg{C}_{12}/\varepsilon$ looks singular and may not have a smooth
limit in $\varepsilon\to 0$ at the first sight.
This problem is solved as follows.
We shall evaluate this term on the representation of
$\alg{d}(2,1;\varepsilon)$, and expand the result in $\varepsilon$,
\begin{align}
\dfrac{1}{\varepsilon}\cT^\alg{C}_{12}
={\mathcal O}(1/\varepsilon)
+{\mathcal O}(1)+{\mathcal O}(\varepsilon)+\cdots~.
\label{sing}
\end{align}
Then, we will find that the most singular term is constant which is
independent of the states in the representation space.
This means that we can interpret it as an overall phase factor of the
S-matrix, which is not relevant in our present analysis.
On the other hand, the higher terms of ${\mathcal O}(\varepsilon)$ simply
vanish after we take the limit $\varepsilon\to 0$.
Therefore the only relevant contribution is the ${\mathcal O}(1)$ term.
We shall see that this term reproduces the non-canonical terms of the
$\alg{su}(2|2)$ classical r-matrix \eqref{bsr}.

\subsection{Representation of $\alg{d}(2,1;\varepsilon)$}
Of course, it is better if we can show that the generator $\fI$ is
the $\varepsilon$-correction of $\fC^\alg{a}{}_\alg{b}$ at the algebraic
level without referring to the representation.
However, we cannot find so far a rigorous argument for this statement.
For this reason, let us construct the representation of
$\alg{d}(2,1;\varepsilon)$ first and evaluate the contribution of the
${\mathcal O}(1)$ term in \eqref{sing}.

In the case of $\alg{psu}(2|2)\ltimes{\mathbb R}^3$, the fundamental
representation $\bs{2}|\bs{2}$ consists of two bosons and two fermions.
It is surprising that in the case of $\alg{d}(2,1;\varepsilon)$ there is
a representation, tantalizingly similar to this fundamental
representation $\bs{2}|\bs{2}$ \cite{VDJ}\footnote{We are grateful to
A.~Torrielli for valuable discussions on this reference.
See also \cite{S}.}.
The only difference is that the root lattice of $\alg{d}(2,1;\varepsilon)$
has one additional dimension, which requires the state to have an
additional index $n$ labeling the weight in this dimension
(related to the grading).
Here, $\fP$ and $\fK$ raises and lowers the weight by one unit, while
$\fQ^\al{}_a$ and $\fS^a{}_\al$ raises and lowers by half of it.
Therefore, it is natural to assume that the representation of the
fermionic generators takes the following indices\footnote{In this
convention, the indices of bosons are integers while those of fermions
are half-integers.
Also, the index of each coefficient coincides with that of the boson.
Though this convention makes the following calculation simple, of course
the final results do not depend on the convention.}: $(n\in{\mathbb Z})$
\begin{align}
&{\fQ}^\alpha{}_a|\phi_n^b\rangle
=a_n{\,}\delta^b_a|\psi_{n+\frac{1}{2}}^\alpha\rangle~,\quad
{\fQ}^\alpha{}_a|\psi_{n-\frac{1}{2}}^\beta\rangle
=b_{n}{\,}\epsilon^{\alpha\beta}\epsilon_{ab}|\phi_{n}^b\rangle~,
\nonumber\\
&{\fS}^a{}_\alpha|\phi_n^b\rangle
=c_{n}{\,}\epsilon^{ab}\epsilon_{\alpha\beta}
|\psi_{n-\frac{1}{2}}^\beta\rangle~,\quad
{\fS}^a{}_\alpha|\psi_{n+\frac{1}{2}}^\beta\rangle
=d_n{\,}\delta^\beta_\alpha|\phi_n^a\rangle~. 
\end{align}
Note that the index $n$ should not be confused with the previous one
$k$ in the representation of $\alg{psu}(2|2)\ltimes{\mathbb R}^3$
\eqref{QSphipsi}, which stands for the $k$-th excitation in the spin
chain.
For the consistency with the commutation relations, especially
\eqref{CRsu}, we have to impose the following conditions,
\begin{align}
1=a_nd_n-b_{n}c_{n}~,\quad
1-\varepsilon=a_nd_n-b_{n+1}c_{n+1}~,
\label{ad-bc_n}
\end{align}
as well as determine the action of bosonic generators as follows:
\begin{align}
&\fP|\phi_{n}^a\rangle 
=a_{n}b_{n+1}|\phi^a_{n+1}\rangle~,\quad
\fP|\psi_{n+\frac{1}{2}}^\alpha\rangle 
=a_{n+1}b_{n+1}|\psi^\alpha_{n+\frac{3}{2}}\rangle~,\nonumber\\
&\fC|\phi_{n}^a\rangle
=\frac{1}{2}(a_{n}d_{n}+b_{n}c_{n})|\phi^a_{n}\rangle~,\quad
\fC|\psi_{n+\frac{1}{2}}^\alpha\rangle 
=\frac{1}{2}(a_{n}d_{n}+b_{n+1}c_{n+1})
|\psi^\alpha_{n+\frac{1}{2}}\rangle~,\nonumber\\
&\fK|\phi_{n}^a\rangle
=c_{n}d_{n-1}|\phi^a_{n-1}\rangle~,\quad
\fK|\psi_{n+\frac{1}{2}}^\alpha\rangle 
=c_{n}d_{n}|\psi^\alpha_{n-\frac{1}{2}}\rangle~.
\end{align}
Note that the indices in the above actions of $\fC$, $\fP$ and $\fK$ are
slightly different between bosons and fermions.
This fact already implies the appearance of the generator $\fI$.

Comparing this representation with the previous $\alg{su}(2|2)$ spin
chain, we can assign a physical interpretation to the current
representation.
Since the grading in the $\alg{su}(2|2)$ spin chain denotes insertion or
removal of the vacuum field ${\mathcal Z}$, it is not difficult to
imagine that the index $n$ can be interpreted as the position of the
excitation on the spin chain.
This interpretation will be helpful in reproducing the non-canonical
terms of the AdS/CFT classical r-matrix later.

The reader may wonder how to reproduce the AdS/CFT classical r-matrix
without the position index from the representation of the exceptional
superalgebra $\alg{d}(2,1;\varepsilon)$ with the position index $n$.
A naive guess is to sum over the states, which makes states blind to the
position index.
Actually this idea is realized in some sense\footnote{We are grateful to
H.~Kanno for discussions on this point.}.
Speaking in a more sophisticated way, we will diagonalize the action of
$(1/\varepsilon)\cT^\alg{C}_{12}$ by Fourier-transforming the above
coordinate (site) picture into the momentum picture as in the asymptotic
state \eqref{asympt}.

{}From the first constraint in \eqref{ad-bc_n}, we can solve $a_n$,
$b_n$, $c_n$ and $d_n$ by parameters $\gamma_n$ and $x_n$ and a constant
$\alpha$:
\begin{align}
a_n=\sqrt{g}\gamma_n~,\quad
b_n=\frac{\alpha}{i\sqrt{g}\gamma_n}D_n~,\quad
c_n=\frac{i\sqrt{g}\gamma_n}{\alpha}x_n^{-1}~,\quad
d_n=\frac{1}{\sqrt{g}\gamma_n}x_nD_n~,
\end{align}
where $D_n=1/(x_n-x_n^{-1})$ and $\sqrt{g}\gamma_n$ scales as a
constant in the limit $g\to\infty$.
This expression is, of course, inspired by the $\alg{su}(2|2)$ spin
chain \eqref{abcd}.
Here we have also exploited the physical input that $\alpha$ is
constant.
The second constraints in \eqref{ad-bc_n} implies
\begin{align}
\frac{x_{n+1}^{-1}}{x_{n+1}-x_{n+1}^{-1}}
-\frac{x_{n}^{-1}}{x_{n}-x_{n}^{-1}}=\varepsilon~,
\end{align}
which means the difference of $x_n$ is of ${\mathcal O}(\varepsilon)$.
If we define $\delta x_n=x_{n+1}-x_n$, the above relation at 
${\mathcal O}(\varepsilon)$ reduces to
\begin{align}
\delta x_n=-\frac{\varepsilon}{2}x_nD_n^{-2}~.
\end{align}

As we will see later, the effect of ${\mathcal O}(\varepsilon)$ will
cancel the singular coefficient $1/\varepsilon$ of the Casimir operator
$\cT^\alg{C}_{12}$ in \eqref{d12} and give a finite contribution.
Therefore, the difference in the indices is very important.
For this reason, although originally fermions have the indices of
half-integers, if we want to compare fermions with bosons, we have to
interpolate the indices into integers\footnote{The important thing is
to balance the bosonic indices with the fermionic ones.
We can alternatively interpolate the bosonic indices into
half-integers as another convention.}.
Note that fermions with integral indices do not exist in the
representation space.
We only introduce these states virtually for the comparison between
bosons and fermions.
(See Fig.~3.)

\begin{figure}[t]
\begin{center}
\scalebox{1.0}[1.0]{\includegraphics{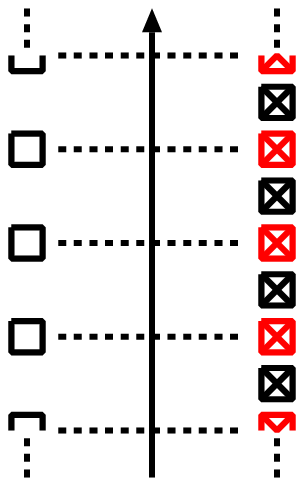}}
\setlength{\unitlength}{1mm}
\begin{picture}(20,20)
\put(-45,20){\makebox(10,10){bosons}}
\put(2,25){\makebox(10,10){fermions}}
\put(-19,50){$\fC$}
\end{picture}
\end{center}
\caption{Virtual fermionic states with integer indices (Red).
The other two dimensions of the weight lattice in Fig. 2 are omitted.}
\end{figure}

After interpolating the index of fermionic states into integers, the
action of the bosonic generators reduces to
\begin{align}
&\fP|\phi_{n}^a\rangle
=\frac{\alpha}{i}
\Bigl[1-\frac{\delta\gamma_n}{\gamma_n}\Bigr]
\Bigl(D_n+\frac{\varepsilon}{2}(x_n+x_n^{-1})\Bigr)
|\phi_{n+1}^a\rangle~,\quad
\fP|\psi_{n}^\alpha\rangle 
=\frac{\alpha}{i}
\left(D_n+\frac{\varepsilon}{4}(x_n+x_n^{-1})\right)
|\psi^\alpha_{n+1}\rangle~,\nonumber\\
&\fC|\phi_{n}^a\rangle
=\frac{1}{2}\bigl(x_n+x_n^{-1}\bigr)D_n|\phi^a_{n}\rangle~,\quad
\fC|\psi_{n}^\alpha\rangle
=\frac{1}{2}\bigl(x_n+x_n^{-1}\bigr)D_n|\psi^\alpha_{n}\rangle~,\nonumber\\
&\fK|\phi_{n}^a\rangle
=\frac{i}{\alpha}
\Bigl[1+\frac{\delta\gamma_n}{\gamma_n}\Bigr]
\Bigl(D_n-\varepsilon x_n^{-1}\Bigr)|\phi_{n-1}^a\rangle~,\quad
\fK|\psi_{n}^\alpha\rangle=\frac{i}{\alpha}
\Bigl(D_n-\frac{\varepsilon}{4}(x_n+x_n^{-1})\Bigr)
|\psi^\alpha_{n-1}\rangle~,
\end{align}
where we have defined $\delta\gamma_n=\gamma_{n+1}-\gamma_n$.

If we take the unitarity condition \cite{Banalytic} into account, the
variable $\gamma_n$ is related to $x_n$ by
\begin{align}
\sqrt{g}\gamma_n=\sqrt{x_nD_n}~,
\label{unitary}
\end{align}
which means $\delta\gamma_n/\gamma_n=\varepsilon x_n^{-1}D_n^{-1}/2$.
In this case, the action of $\fP$ and $\fK$ is slightly simplified:
\begin{align}
\fP|\phi_{n}^a\rangle
=\frac{\alpha}{i}\left(D_n+\frac{\varepsilon}{2}x_n\right)
|\phi^a_{n+1}\rangle~,\quad
\fK|\phi_{n}^a\rangle
=\frac{i}{\alpha}\left(D_n-\frac{\varepsilon}{2}x_n^{-1}\right)
|\phi^a_{n-1}\rangle~.
\end{align}

\subsection{Classical r-matrix}
After we constructed the representation in the previous subsection, let
us reproduce the non-canonical terms of the classical r-matrix
\eqref{bsr}.
We evaluate the singular term \eqref{sing} explicitly on a two-site state
$|\chi_n\chi'_m\rangle$, defined by a tensor product
$|\chi_n\rangle_1\otimes|\chi'_m\rangle_2$:
\begin{align}
\frac{1}{\varepsilon}\cT^\alg{C}_{12}|\chi_n\chi'_m\rangle
=r_n|\chi_{n+1}\chi'_{m-1}\rangle
+s_n|\chi_n\chi'_m\rangle
+t_n|\chi_{n-1}\chi'_{m+1}\rangle~,
\label{eigenprob}
\end{align}
with $r_n$, $s_n$ and $t_n$ being some appropriate coefficients.
In general, we find the action of $\cT^\alg{C}_{12}$ generates a linear
combination of the states with different position indices.
Comparing with the asymptotic state \eqref{asympt} which is a momentum
eigenstate, it is not difficult to convince ourselves that we have to
consider the eigenstate of $\cT^\alg{C}_{12}$.
Therefore, let us solve the infinite-dimensional eigenvalue problem.
Note that the reality condition of the eigenvalue is guaranteed only
when the infinite-dimensional matrix \eqref{eigenprob} is symmetric.
This is the case if we impose the unitarity condition \eqref{unitary}.

Since generally a matrix has plural eigenvalues, we have to specify
which eigenvalue or eigenstate we want to perturb from.
In the classical limit $g\to\infty$ of the $\alg{su}(2|2)$ spin chain,
\eqref{xpmsol} gives $x^+=x^-$ which implies the excitation has the zero
momentum $p=0$ from \eqref{U}.
Hence, the asymptotic state \eqref{asympt} is simply the summation of
states with excitations at each site with weight one.
Motivated by this $\alg{su}(2|2)$ spin chain, we are led to
studying the perturbation theory from the eigenstate summing up all the
states with weight one,
\begin{align}
\sum_{l=-\infty}^\infty|\chi_{n-l}\chi'_{m+l}\rangle~.
\end{align}

The solution to this eigenvalue problem is well-established in the
perturbation theory of quantum mechanics \cite{QM}.
In a system with Hamiltonian $H^{(0)}+\varepsilon H^{(1)}$ where
$H^{(0)}$ is solved exactly to have the eigenvalue $E^{(0)}_\ell$ with
the eigenstate $|\varphi^{(0)}_\ell\rangle$, the first-order deviation
of the eigenvalue $E^{(1)}_\ell$ is given by a famous formula,
\begin{align}
E^{(1)}_\ell
=\langle\varphi^{(0)}_\ell|H^{(1)}|\varphi^{(0)}_\ell\rangle~.
\label{qm}
\end{align}

In our present case where the eigenstate before the perturbation is a
summation of weight one, the first-order deviation is simply given by
summing up the matrix elements indiscriminately with weight one:
\begin{align}
\frac{1}{\varepsilon}\cT^\alg{C}_{12}|\chi_n\chi'_m\rangle
\sim(r_n+s_n+t_n)|\chi_n\chi'_m\rangle~.
\end{align}
After the calculation, we find (independently of $\delta\gamma/\gamma$)
\begin{align}
\frac{1}{\varepsilon}\cT^\alg{C}_{12}|\phi^a_1\phi^b_2\rangle
&=\Bigl[\frac{1}{\varepsilon}T
-\frac{1}{2}\left(D^{-1}_1D_2+D_1D^{-1}_2\right)\Bigr]
|\phi^a_1\phi^b_2\rangle~,\nonumber\\
\frac{1}{\varepsilon}\cT^\alg{C}_{12}|\psi^\al_1\psi^\be_2\rangle
&=\Bigl[\frac{1}{\varepsilon}T+0\Bigr]
|\psi^\al_1\psi^\be_2\rangle~,\nonumber\\
\frac{1}{\varepsilon}\cT^\alg{C}_{12}
|\phi^a_1\psi^\al_2\rangle
&=\Bigl[\frac{1}{\varepsilon}T-\frac{1}{2}D^{-1}_1D_2\Bigr]
|\phi^a_1\psi^\al_2\rangle~,\nonumber\\
\frac{1}{\varepsilon}\cT^\alg{C}_{12}|\psi^\al_1\phi^a_2\rangle
&=\Bigl[\frac{1}{\varepsilon}T-\frac{1}{2}D_1D^{-1}_2\Bigr]
|\psi^\al_1\phi^a_2\rangle~,
\end{align}
where $T$ is a quantity of ${\mathcal O}(1)$,
\begin{align}
T=\dfrac{1}{2}(x_1+x^{-1}_1)(x_2+x^{-1}_2)D_1D_2-2D_1D_2~,
\label{T}
\end{align}
and we have dropped higher order terms of $\varepsilon$.
Since the variables associated to the $1$-st and $2$-nd excitations
always have the position indices $n$ and $m$, here we have used the
``wavepacket'' indices $1$ and $2$ instead of the position indices.
By subtracting the average of the right-hand-sides, we can normalize the
classical r-matrix as
\begin{align}
\frac{1}{\varepsilon}\cT^\alg{C}_{12}|\phi^a_1\phi^b_2\rangle
&=\Bigl[-\frac{1}{4}D^{-1}_1D_2-\frac{1}{4}D_1D^{-1}_2\Bigr]
|\phi^a_1\phi^b_2\rangle~,\nonumber\\
\frac{1}{\varepsilon}\cT^\alg{C}_{12}|\psi^\al_1\psi^\be_2\rangle
&=\Bigl[+\frac{1}{4}D^{-1}_1D_2+\frac{1}{4}D_1D^{-1}_2\Bigr]
|\psi^\al_1\psi^\be_2\rangle~,\nonumber\\
\frac{1}{\varepsilon}\cT^\alg{C}_{12}|\phi^a_1\psi^\al_2\rangle
&=\Bigl[-\frac{1}{4}D^{-1}_1D_2+\frac{1}{4}D_1D^{-1}_2\Bigr]
|\phi^a_1\psi^\al_2\rangle~,\nonumber\\
\frac{1}{\varepsilon}\cT^\alg{C}_{12}|\psi^\al_1\phi^a_2\rangle
&=\Bigl[+\frac{1}{4}D^{-1}_1D_2-\frac{1}{4}D_1D^{-1}_2\Bigr]
|\psi^\al_1\phi^a_2\rangle~.
\end{align}
This is nothing but the non-canonical terms of the AdS/CFT classical
r-matrix \eqref{bsr}.

Since we are solving an infinite-dimensional eigenvalue problem, it is
safer to present the eigenstates as well, instead of simply applying the
formula \eqref{qm} from the perturbation theory.
For this purpose we have to know that $r_n$, $s_n$ and $t_n$ appeared
above satisfy
\begin{align}
s_{n+1}-s_n=-u_1D_1+u_2D_2~,\quad
r_{n+1}-r_n=t_{n+1}-t_n=\frac{1}{2}(u_2D_1-u_1D_2)~,
\end{align}
with the spectral parameter $u=x+x^{-1}$.
This recursive relation is independent of the states in the
representation space, because it comes from the universal singular term
$T$ \eqref{T}.
If we assume the eigenstate to be
\begin{align}
|\chi_1\chi'_2\rangle=\sum_{l=-\infty}^\infty
(1+\varepsilon f_{n-l})|\chi_{n-l}\chi'_{m+l}\rangle~,
\end{align}
we will find a recursive relation for $f_n$ independent of bosons or
fermions:
\begin{align}
f_{n+1}-2f_n+f_{n-1}=f_1-2f_0+f_{-1}+n(u_2-u_1)(D_1^{-1}+D_2^{-1})~.
\end{align}
The recursive relation can be solved by
\begin{align}
f_n=\frac{n^3}{6}(u_2-u_1)(D_1^{-1}+D_2^{-1})~,
\end{align}
with suitable initial conditions.
It is easy to reproduce the above eigenvalues from this eigenvector.
We thus complete our solution to the eigenvalue problem.

\section{Conclusions}
In this paper, we have investigated the AdS/CFT spin chain with the
symmetry $\alg{psu}(2|2)\ltimes{\mathbb R}^3$ using the exceptional Lie
superalgebra $\alg{d}(2,1;\varepsilon)$.
In our analysis, we have obtained two results.
First, we have rederived the coproducts of the level-1 Yangian
generators from $\alg{d}(2,1;\varepsilon)$.
In the derivation, we find that the non-trivial part of the secret
symmetry $\fI$ appears as the $\varepsilon$-correction of the last
$\alg{su}(2)$ triplet generators $\fC^\alg{a}{}_\alg{b}$ in
$\alg{d}(2,1;\varepsilon)$.
Secondly we have reproduced the non-canonical AdS/CFT classical r-matrix
\eqref{bs} from the canonical r-matrix of $\alg{d}(2,1;\varepsilon)$. 
The non-canonical terms of the classical r-matrix come from the
differences of the action of $\fC^\alg{a}{}_\alg{b}$ on bosons and
fermions.

Originally, as we explained in the introduction, three regularizations
were adopted to cure the degeneracy of the Killing form: superalgebra
$\alg{d}(2,1;\varepsilon)$, $\alg{su}(2)$ outer automorphism and
superalgebra $\alg{u}(2|2)$.
The $\alg{su}(2)$ outer automorphism does not look intrinsic to the
$\alg{su}(2|2)$ spin chain model.
In our analysis, we have given an interpretation to all these
regularizations in $\alg{d}(2,1;\varepsilon)$ and made the outer
$\alg{su}(2)$ automorphism look more intrinsic to the model.

Let us conclude by listing several further directions.
\begin{itemize}
\item
The super Yang-Mills theory does not have the symmetry
$\alg{d}(2,1;\varepsilon)$.
At this stage, pursuing the Yangian symmetries in
$\alg{d}(2,1;\varepsilon)$ is a purely technical tool to access
various results of the Yangian algebra with the non-degenerating
Killing form and to study the secret symmetry as the
$\varepsilon$-correction.
However, considering the success of the off-shell formalism with the
centrally extended $\alg{su}(2|2)$ algebra and the naturalness of
reproduction of this algebra from $\alg{d}(2,1;\varepsilon)$ in the
limit $\varepsilon\to 0$, it may be eventually possible to assign a
physical meaning to the exceptional superalgebra
$\alg{d}(2,1;\varepsilon)$ in the super Yang-Mills theory (or its
deformation).
\item 
Although our derivation of the AdS/CFT classical r-matrix from the
      exceptional superalgebra $\alg{d}(2,1;\varepsilon)$ fully makes
      sense from the viewpoint of the representation theory, it is not
      easy to assign a physical interpretation to the
      $\alg{d}(2,1;\varepsilon)$ spin chain.
Especially, it is not very clear whether the mathematical tensor product
      of the single-excitation states really corresponds to the physical
      picture of a spin chain state with multi-excitations.
Since the representation of $\alg{d}(2,1;\varepsilon)$ has the position
      index, making full sense of the $\alg{d}(2,1;\varepsilon)$ spin
      chain may be helpful in understanding the finite size effects
      \cite{finite}.
Also, since we have subtracted the overall shift to derive the
      non-canonical expression of the classical r-matrix, we expect the
      $\alg{d}(2,1;\varepsilon)$ spin chain gives an implication to the
      overall phase factor \cite{phase}.
\item 
In section 3 we have reproduced the non-trivial coproduct of the
      symmetry $\fI$ as the $\varepsilon$-correction of the generator
      $\fC$, while the non-canonical term of the AdS/CFT classical
      r-matrix seems to be reproduced directly from the
      $\varepsilon$-correction of the generators $\fP$ and $\fK$ in
      section 4.
We believe that the difference is ``convention-dependent'', though we
      cannot make this statement more clear.
A similar question whether we can construct a secret symmetry with the
      non-trivial coproduct being
$\epsilon^{ab}\epsilon_{\alpha\beta}
\fQ^\alpha{}_a\cU^{+1}\otimes\fQ^\beta{}_b$ 
or
$\epsilon^{\alpha\beta}\epsilon_{ab}
\fS^a{}_\alpha\cU^{-1}\otimes\fS^b{}_\beta$ \cite{BS}
remains unanswered.
\item 
Though various deformations are introduced to investigate the
      $\alg{su}(2|2)$ spin chain \cite{BK}, our work suggests that we
      can lift the question of the universal R-matrix of the AdS/CFT
      spin chain into that of the exceptional superalgebra
      $\alg{d}(2,1;\varepsilon)$.
Work on the universal R-matrix of $\alg{d}(2,1;\varepsilon)$ \cite{HSTY}
      seems promising in finding the universal R-matrix of the AdS/CFT
      spin chain.
\item
As we have mentioned in the introduction, two possible viewpoints can
be assigned to the secret symmetry $\fI$: as the $\alg{su}(2)$
automorphism and as the composite operator $\cT^\alg{psu}\fC^{-1}$.
In lifting the AdS/CFT spin chain to the $\alg{d}(2,1;\varepsilon)$ model,
      we believe we have given both $\fI$ and $\fB^\alg{a}{}_\alg{b}$ a
      nice picture as the $\varepsilon$-correction of the last
      $\alg{su}(2)$ generators $\fC^\alg{a}{}_\alg{b}$.
However, it is still unclear how the interpretation as a composite
      operator is consistent from the $\alg{d}(2,1;\varepsilon)$
      viewpoint.
Understanding the meaning of the composite operator interpretation may
      help us in deriving the non-canonical classical r-matrix from
      $\alg{d}(2,1;\varepsilon)$ at the algebraic level without
      mentioning to the representation.
\item
It is always interesting to study various aspects of this spin chain
      model from the string worldsheet theory.
Recent works \cite{worldsheet} may give a clue in this direction.
\end{itemize}

\section*{Acknowledgement}
We are grateful to H.~Awata, H.Y.~Chen, A.~Hashimoto, H.~Hayashi,
N.~Ishibashi, Y.~Satoh, S.~Teraguchi, Y.~Yonezawa, especially our
colleague H.~Kanno and our previous collaborator A.~Torrielli for
valuable discussions.
Part of this work was done during ``The 2nd Asian Winter School on
String Theory'' at Kusatsu in Japan.
We would also like to thank the lecturers S.~Minwalla and M.~Staudacher
for patiently answering our basic questions related to the subject of
this paper.
We are also grateful to N.~Beisert and F.~Spill for their valuable
comments on the previous version of this paper.
The work of S.M. is supported partly by Inamori Foundation and partly by
Grant-in-Aid for Young Scientists (B) [\#18740143] from the Japan
Ministry of Education, Culture, Sports, Science and Technology.


\begin{thebibliography}{99}
\bibitem{MZ}
J.~A.~Minahan and K.~Zarembo,
``The Bethe-ansatz for N = 4 super Yang-Mills,''
JHEP {\bf 0303}, 013 (2003)
[arXiv:hep-th/0212208].

\bibitem{BPR}
I.~Bena, J.~Polchinski and R.~Roiban,
``Hidden symmetries of the AdS(5) x S**5 superstring,''
Phys.\ Rev.\  D {\bf 69}, 046002 (2004)
[arXiv:hep-th/0305116].

\bibitem{review}
N.~Beisert,
``The dilatation operator of N = 4 super Yang-Mills theory and
integrability,''
Phys.\ Rept.\  {\bf 405}, 1 (2005)
[arXiv:hep-th/0407277].
A.~A.~Tseytlin,
``Semiclassical strings and AdS/CFT,''
arXiv:hep-th/0409296.
J.~Plefka,
``Spinning strings and integrable spin chains in the AdS/CFT
correspondence,''
Living Rev.\ Rel.\  {\bf 8}, 9 (2005)
[arXiv:hep-th/0507136].
J.~A.~Minahan,
``A Brief Introduction To The Bethe Ansatz In N=4 Super-Yang-Mills,''
J.\ Phys.\ A  {\bf 39}, 12657 (2006).

\bibitem{BSmatrix}
N.~Beisert,
``The $su(2|2)$ dynamic S-matrix,''
arXiv:hep-th/0511082.

\bibitem{BD}
A.~A.~Belavin and V.~G.~Drinfel'd,
``Solutions of the classical Yang-Baxter equation for simple Lie
algebras,''
Funct.\ Anal.\ Appl.\ {\bf 16} 159 (1982). 

\bibitem{M}
Z.~Q.~Ma,
``Yang-Baxter equation and quantum enveloping algebras,''
Advanced series on theoretical physical science, 
World Scientific (1993).
N.~J.~MacKay,
``Introduction to Yangian symmetry in integrable field theory,''
Int.\ J.\ Mod.\ Phys.\  A {\bf 20}, 7189 (2005)
[arXiv:hep-th/0409183].

\bibitem{Banalytic}
N.~Beisert,
``The Analytic Bethe Ansatz for a Chain with Centrally Extended
$su(2|2)$ Symmetry,''
J.\ Stat.\ Mech.\  {\bf 0701}, P017 (2007)
[arXiv:nlin/0610017].

\bibitem{MT}
S.~Moriyama and A.~Torrielli,
``A Yangian Double for the AdS/CFT Classical r-matrix,''
JHEP {\bf 0706}, 083 (2007)
[arXiv:0706.0884 [hep-th]].

\bibitem{MMT}
T.~Matsumoto, S.~Moriyama and A.~Torrielli,
``A Secret Symmetry of the AdS/CFT S-matrix,''
JHEP {\bf 0709}, 099 (2007)
[arXiv:0708.1285 [hep-th]].

\bibitem{BS}
N.~Beisert and F.~Spill,
``The Classical r-matrix of AdS/CFT and its Lie Bialgebra Structure,''
arXiv:0708.1762 [hep-th].

\bibitem{Hopf}
C.~Gomez and R.~Hernandez,
``The magnon kinematics of the AdS/CFT correspondence,''
JHEP {\bf 0611}, 021 (2006)
[arXiv:hep-th/0608029].
J.~Plefka, F.~Spill and A.~Torrielli,
``On the Hopf algebra structure of the AdS/CFT S-matrix,''
Phys.\ Rev.\  D {\bf 74}, 066008 (2006)
[arXiv:hep-th/0608038].

\bibitem{AF}
G.~Arutyunov and S.~Frolov,
``On AdS(5) x S**5 string S-matrix,''
Phys.\ Lett.\  B {\bf 639}, 378 (2006)
[arXiv:hep-th/0604043].

\bibitem{T}
A.~Torrielli,
``Classical r-matrix of the $su(2|2)$ SYM spin-chain,''
Phys.\ Rev.\  D {\bf 75}, 105020 (2007)
[arXiv:hep-th/0701281].

\bibitem{BYangian}
N.~Beisert,
``The S-Matrix of AdS/CFT and Yangian Symmetry,''
PoS {\bf SOLVAY}, 002 (2006)
[arXiv:0704.0400 [nlin.SI]].

\bibitem{VDJ}
J.~Van Der Jeugt,
``Irreducible Representations Of The Exceptional Lie Superalgebras
$D(2,1;\alpha)$,'' 
J.\ Math.\ Phys.\  {\bf 26}, 913 (1985).

\bibitem{S}
F.~Spill, ``Hopf Algebras in the AdS/CFT Correspondence'',
Diploma Thesis, Humboldt University of Berlin, 2007.

\bibitem{QM}
J.~J.~Sakurai, ``Modern Quantum Mechanics'', Addison-Wesley Publishing
Company (1993).
K.~Igi, H.~Kawai, ``Quantum Mechanics'', Kodansha (1994), in Japanese.

\bibitem{finite}
J.~Ambjorn, R.~A.~Janik and C.~Kristjansen,
``Wrapping interactions and a new source of corrections to the
spin-chain / string duality,''
Nucl.\ Phys.\  B {\bf 736}, 288 (2006)
[arXiv:hep-th/0510171].
G.~Arutyunov, S.~Frolov and M.~Zamaklar,
``Finite-size effects from giant magnons,''
Nucl.\ Phys.\  B {\bf 778}, 1 (2007)
[arXiv:hep-th/0606126].
R.~A.~Janik and T.~Lukowski,
``Wrapping interactions at strong coupling -- the giant magnon,''
Phys.\ Rev.\  D {\bf 76}, 126008 (2007)
[arXiv:0708.2208 [hep-th]].
G.~Arutyunov and S.~Frolov,
``On String S-matrix, Bound States and TBA,''
JHEP {\bf 0712}, 024 (2007)
[arXiv:0710.1568 [hep-th]].
Y.~Hatsuda and R.~Suzuki,
``Finite-Size Effects for Dyonic Giant Magnons,''
arXiv:0801.0747 [hep-th].
J.~A.~Minahan and O.~Ohlsson Sax,
``Finite size effects for giant magnons on physical strings,''
arXiv:0801.2064 [hep-th].

\bibitem{phase}
R.~A.~Janik,
``The AdS(5) x S**5 superstring worldsheet S-matrix and crossing
symmetry,''
Phys.\ Rev.\  D {\bf 73}, 086006 (2006)
[arXiv:hep-th/0603038].
N.~Beisert, R.~Hernandez and E.~Lopez,
``A crossing-symmetric phase for AdS(5) x S**5 strings,''
JHEP {\bf 0611}, 070 (2006)
[arXiv:hep-th/0609044].
N.~Beisert, B.~Eden and M.~Staudacher,
``Transcendentality and crossing,''
J.\ Stat.\ Mech.\  {\bf 0701}, P021 (2007)
[arXiv:hep-th/0610251].
K.~Sakai and Y.~Satoh,
``Microscopic formulation of the S-matrix in AdS/CFT,''
JHEP {\bf 0712}, 044 (2007)
[arXiv:0709.3342 [hep-th]].

\bibitem{BK}
N.~Beisert and P.~Koroteev,
``Quantum Deformations of the One-Dimensional Hubbard Model,''
arXiv:0802.0777 [hep-th].

\bibitem{HSTY}
I.~Heckenberger, F.~Spill, A.~Torrielli and H.~Yamane,
``Drinfeld second realization of the quantum affine superalgebras of
$D^{(1)}(2,1:x)$ via the Weyl groupoid,''
arXiv:0705.1071 [math.QA].

\bibitem{worldsheet}
G.~Arutyunov, S.~Frolov, J.~Plefka and M.~Zamaklar,
``The off-shell symmetry algebra of the light-cone AdS(5) x S**5
superstring,''
J.\ Phys.\ A  {\bf 40}, 3583 (2007)
[arXiv:hep-th/0609157].
T.~Klose, T.~McLoughlin, R.~Roiban and K.~Zarembo,
``Worldsheet scattering in AdS(5) x S**5,''
JHEP {\bf 0703}, 094 (2007)
[arXiv:hep-th/0611169].
A.~Mikhailov and S.~Schafer-Nameki,
``Algebra of transfer-matrices and Yang-Baxter equations on the string
worldsheet in AdS(5) x S(5),''
arXiv:0712.4278 [hep-th].
J.~M.~Evans and J.~O.~Madsen,
``Quantum integrability of coupled N = 1 super sine/sinh-Gordon theories
and the Lie superalgebra $D(2,1,\alpha)$,''
Int.\ J.\ Mod.\ Phys.\  A {\bf 14}, 2551 (1999)
[arXiv:hep-th/9712227].

\end{thebibliography}
\end{document}